\documentclass[twocolumn,english,aps,jcp,superscriptaddress,showpacs,floatfix]{revtex4} 
\usepackage{graphicx}
\usepackage{amsmath,amssymb,amsfonts}
\usepackage{natbib}
\usepackage{babel}
\usepackage{lipsum}
\usepackage{tikz}
\usepackage{changes}
\usepackage[shortlabels]{enumitem}
\usetikzlibrary{shapes,arrows}
\usepackage{dcolumn}
\usepackage{bm}

\newcommand{\sref}[1]{Sec. \ref{#1}}

\newcommand{\eref}[1]{Eq.\hspace{0.025in}(\ref{#1})}

\newcommand{\Eref}[1]{Equation (\ref{#1})}

\DeclareMathAlphabet\mathbfcal{OMS}{cmsy}{b}{n}

\begin{document}

\preprint{APS/123-QED}

\title{Identifying weak values with intrinsic dynamical properties in Modal theories}
\author{Devashish Pandey}
\affiliation{Department of Electronic Engineering, Universitat Aut\`onoma de Barcelona, 08193 Bellaterra, Spain}%
\affiliation{Department of Photonics Engineering, Technical University of Denmark, DK-2800 Kgs. Lyngby, Denmark}
\author{Rui Sampaio}
\affiliation{QTF Centre of Excellence, Department of Applied Physics, Aalto University, FI-00076, Aalto, Espoo, Finland}
\author{Tapio Ala-Nissila}
\affiliation{QTF Centre of Excellence, Department of Applied Physics, Aalto University, FI-00076, Aalto, Espoo, Finland}
\affiliation{Interdisciplinary Centre for Mathematical Modelling and Department of Mathematical Sciences, Loughborough University, Loughborough, Leicestershire LE11 3TU, United Kingdom}
\author{Guillermo Albareda}
\affiliation{Max Planck Institute for the Structure and Dynamics of Matter, 22761 Hamburg, Germany}%
\affiliation{Institute of Theoretical and Computational Chemistry, Universitat de Barcelona, 08028 Barcelona, Spain}
\author{Xavier Oriols}
\email{xavier.oriols@uab.cat}
\affiliation{Department of Electronic Engineering, Universitat Aut\`onoma de Barcelona, 08193 Bellaterra, Spain}%


\date{\today}

\begin{abstract}
The so-called eigenvalue-eigenstate link states that no property can be associated to a quantum system unless it is in an eigenstate of the corresponding operator. This precludes the assignation of properties to unmeasured quantum systems in general. 
This arbitrary limitation of Orthodox quantum mechanics generates many puzzling situations such as for example the impossibility to uniquely define a work distribution, an essential building block of quantum thermodynamics.  
Alternatively, Modal theories (e.g., Bohmian mechanics) provide an ontology that always allows to define intrinsic properties, i.e., properties of quantum systems that are detached from any possible measuring context. 
We prove here that Aharonov, Albert and Vaidman's notion of weak value can always be identified with an intrinsic dynamical property of a quantum system defined in a certain Modal theory. Furthermore, the fact that weak values are experimentally accessible (as an ensemble average of weak measurements which are post-selected by a strong measurement) strengthens the idea that understanding the intrinsic (unperturbed) dynamics of quantum systems is possible and useful in a given Modal theory. As examples of the physical soundness of these intrinsic properties, we discuss three intrinsic Bohmian properties, viz., the dwell time, the work distribution and the quantum noise at high frequencies.
 
\end{abstract}



\maketitle


\section{Introduction}

\label{sec1}

Properties of classical systems are well defined regardless of whether they are being measured or not. Therefore, evaluating a property of a system at time $t_1$ and correlating the outcome with the value of the same (or another) property at a later time $t_2$ provides an unequivocal way of representing the dynamics of classical systems. In quantum mechanics, however, Bell~\cite{bellcontext} as well as Kochen and Specker~\cite{Kochen1975}, showed that measurements cannot be thought of as simply revealing the underlying (intrinsic) properties of the system in a way that is independent of the context in which the observable is measured. 
The result of correlating the outcome of measuring an observable of a quantum system at time $t_1$ with that at $t_2$ of the same (or another) observable depends, in general, upon the specification of the measuring context needed to obtain these outcomes. 
This characteristic of quantum phenomena is known as contextuality and the unavoidable perturbation that measurements induce on the subsequent evolution of quantum systems is commonly referred to as quantum backaction~\cite{Dieter,maudlin}.

A practical problem associated to quantum backaction and contextuality arises when trying to define any multi-time (i.e., that requires at least two-time measurements) property of a quantum system. 
Consider for example the case of the quantum work distribution. One can think of evaluating it by means of a two-time projective measurement protocol of the energy~\cite{work_TPM1,work_TPM2}. However, the first measurement projects the initial state into an energy eigenstate and hence it prevents the possibility of capturing any coherent evolution of the energy beyond that of a Hamiltonian eigenstate. One could think that avoiding the quantum backaction is possible by simply making the coupling between the system and measuring apparatus very weak, e.g., using the so-called indirect or weak measurements~\cite{opensystem,vaidman1996weak}. 
Unfortunately, it is not possible to obtain (Born) probability distributions of dynamical properties that are independent of the measurement context (see \cite{App_A} for a detailed explanation).
Therefore, whilst a number of alternative protocols have been proposed to alleviate this problem (based on the use of weak and collective measurements~\cite{work66,work67,work54,work70}), the backaction of the measuring apparatus on the measured system is, in all existing protocols, an undesired side effect that yields a list of incompatible definitions of quantum work. This has culminated in a ``no-go'' theorem that states that, in fact, there cannot exist a (super)operator for work that simultaneously satisfies all the physical properties required from a proper definition of work in quantum systems~\cite{work58}.

In this context, a natural question to ask is whether \textit{intrinsic dynamical properties}, defined as: 
\begin{quote}
    \textit{dynamical properties the existence of which is detached from the measuring apparatus and hence that provide information of the unperturbed (backaction-free) quantum dynamics}
\end{quote}
can be defined and possibly measured~\cite{footnote_new1}. 
The postulates of Orthodox quantum mechanics are categorical~\cite{footnote_new2}: the so-called eigenvalue-eigenstate link establishes that no property can be associated to a quantum system unless it is in an eigenstate of the corresponding operator. The eigenvalue-eigenstate link is motivated (but not imposed) by the experimental fact that a projective measurement of a property of a quantum system that is in an eigenstate of the associated operator provides a single outcome (eigen-) value without ambiguity (i.e., a single experiment is enough to determine the property of the system). On the contrary, a projective measurement of a property of a system that is not in an eigenstate of the corresponding operator can yield many different output values (i.e., a single experiment is not enough to determine the property of a system). This only result, which is in accordance with Born's law, is used by Orthodox quantum mechanics to decide about what can be and what cannot be considered a real property of a quantum system.
That is, Orthodox quantum mechanics rejects the reality of an unmeasured object's property unless the object is in an eigenstate of the property operator~\cite{footnote1}. As such, our previous definition of intrinsic dynamical properties has no place in Orthodox quantum mechanics.

On this matter, the purpose of this work is threefold. Our first aim is to establish the correct theoretical framework under which an intrinsic dynamical property is a well defined quantity. Second, we want to answer the question of whether an intrinsic dynamical property can be experimentally accessed. And third, we would like to provide some relevant examples substantiating the importance of these type of properties. Accordingly, we have structured the paper as follows. In \sref{modal} we introduce Modal theories and show that these theories are a valid mathematical framework to define intrinsic dynamical properties. In \sref{connection} we will demonstrate that an intrinsic property and a weak value are the same precise thing from the point of view of Modal quantum mechanics. This will be a proof that intrinsic properties can be actually measured. The implications of this connection between intrinsic properties and weak values will be discussed in \sref{implications}. In \sref{examples_unmeasured}, 
we will acknowledge Bohmian mechanics as a particular Modal theory, and discuss three paradigmatic examples of the soundness of these type of intrinsic dynamical properties: the dwell time, the quantum work distribution, and the quantum noise at high frequencies.  The summary and conclusions of the work will be presented in \sref{conclusion}.

\section{Intrinsic dynamical properties in Modal theories}\label{modal}

A \textit{static property} contains information of a system and an operator at a given time. Alternatively, a \textit{dynamical property} bears information about a system associated to, at least, one operator at two different times or two different (non-commuting) operators at the same time.
Therefore, in Orthodox quantum mechanics it is not possible to define a \textit{real} intrinsic (backaction-free) dynamical property even though, it is well accepted that adopting the Orthodox ontology for giving reality to only measured values is a deliberate choice that is not imposed by any experimental fact~\cite{bong2020strong}.

In this respect, van Fraassen proposed to develop new quantum theories without imposing the ``eigenvalue-eigenstate link''~\cite{van1981modal}. These theories are today known as Modal quantum theories~\cite{van1981modal,bub1996modal,bub1999interpreting}. The main idea is to deny a special status to measurements (measurements should be dealt with in the same way as ordinary physical interactions) so that the usual unitary evolution of the quantum state in Hilbert space has to be valid at all times, with or without measurement. How can the unitary evolution of the quantum state (which may imply a superposition of different eigenstates) be made compatible with the experimental evidence of getting a definite eiegnvalue? The simplest solution is by introducing, apart from the quantum state of the Orthodox theory $|\psi(t)\rangle$ , which we refer to as \textit{guiding state}, an additional quantum state $|g^i(t)\rangle$, named \textit{property states} that specifies the value $g^i(t)$ of the property $G$ for a given $i$-th experiment at time $t$. Property states represent real valued properties of the quantum system that are not defined alongside any measurement process and therefore have no counterpart in the Orthodox theory. While other quantum interpretations are often put under the umbrella of what is commonly known as a Modal theory~\cite{dieks2007probability}, we will here only consider theories dealing with one \textit{property state} added to the \textit{guiding state}~\cite{footnote_new3}. 

The quantum state in a Modal theory is thus defined not only by  $|\psi(t) \rangle$, but also by $g^i(t)$ that is also referred to as the \textit{ontic} variable (as will be seen later, the superindex $i$ refers to a specific experiment). Thus, when the system is not coupled to a measuring apparatus, and hence there is no possible quantum backaction, the value $g^i(t)$ defines an intrinsic property of the quantum system. In this sense, Modal theories open the path for defining \textit{intrinsic dynamical properties} at the ontological level. 
Note that Modal interpretations do not reject the existence of projective measurements. In fact, the outcome of a projective measurement in Orthodox quantum mechanics is identical to the outcome of the same projective measurement performed within any Modal theory (otherwise Orthodox and Modal theories would be empirically distinguishable, which is not the case). When a property $G$ is projectively measured at a given time $t$ (for a given experiment $i$), one obtains the value $g^i(t)$, which obviously will coincide with a particular eigenvalue of the operator $\hat G$.  
Quantum uncertainty is then recovered by an ensemble of $N$ identically prepared experiments that are all associated to the same \textit{guiding state} $|\psi(t) \rangle$ but to different \textit{property states} $|g^i(t)\rangle$. 
Consequently, Modal theories are empirically equivalent to Orthodox quantum mechanics because the probability density distribution of the property $G$ at time $t$ always satisfies Born's law. A mathematical way of expressing this condition is $|\psi(g,t)|^2 = \lim_{N\to\infty} \frac{1}{N}\sum_{i=1}^N \delta(g-g^i(t))$ (where $\delta$ plays the role of a Kronecker or a Dirac delta depending on whether integrals over $g$ are involved or not). Then, the ensemble value of $G$ provided by the Modal theory from different identically-prepared experiments reads:
\begin{equation}
\langle {G}\rangle_M = \lim_{N\rightarrow\infty} \frac{1}{N} \sum_{i=1}^N g^i(t) = \int dg \;g|\psi(g,t)|^2  = \langle {G}\rangle,
\label{modal1} 
\end{equation}
where for the case of a (partially) discrete spectrum the integral should be interpreted as a Stieltjes one. 
That is, \eref{modal1} states that the expectation value of a property $G$ is identical in Modal, $\langle {G}\rangle_M$, and Orthodox quantum mechanics, $\langle \hat{G}\rangle$. In this sense, we say that static intrinsic properties are 'unproblematic' within the Orthodox theory.

Despite the fact that the Orthodox theory can provide expectation values of static properties, strictly speaking, intrinsic static properties are not well defined in Orthodox quantum mechanics. That is, before a projective measurement is carried out, Orthodox quantum mechanics asserts that the quantum system defined by the state $|\psi(t) \rangle$ has no property $G$ associated unless $|\psi(t) \rangle=|g \rangle$~\cite{pandey2020micro}.

\section{Connection between weak values and intrinsic properties}
\label{connection}

Among a great variety of Modal theories, Bohmian mechanics is probably the most prominent example. In Bohmian mechanics the \textit{property state} of the position is related to the so-called Bohmian trajectory, which in turn evolves according to the Bohmian velocity field~\cite{bub1996modal}.
The position and momentum operators do not commute and hence there are no simultaneous eigensates of both properties. The Orthodox theory thus concludes that the velocity at a given position is not a well-defined property for a quantum particle. In 2007, however, Howard Wiseman~\cite{wiseman2007grounding} showed that an operationalist definition of the velocity of a quantum particle involving weak and projective measurements separated by an infinitesimal lapse of time  (``using a technique that would make sense even to a physicist with no knowledge of quantum mechanics'') leads to the concept of weak value as defined by Aharonov, Albert and Vaidman~\cite{aharonov1988result}. This weak value was proven to be the Bohmian velocity~\cite{wiseman2007grounding,durr2009weak}. Following this theoretical finding, a number of experiments have been carried out where the Bohmian velocity  has been ``measured'' in the laboratory~\cite{kocsis2011observing,mahler2016experimental,xiao2017experimental,xiao2019observing,vaidman2017weak}.

But, if the Bohmian velocity can be measured by means of weak values, why is this property not well-defined in Orthodox quantum mechanics? 
Let us answer this apparent contradiction. 
Weak values can be experimentally accessed only by averaging over a sub-ensemble of post-selected two-time measurements. This experimental procedure, which involves a post-processing of the data acquired in an ensemble of experiments, clearly goes beyond the limits of what can be accepted as a real property within the eigenvalue-eigenstate link. We remind that the Orthodox concept of reality is based on our ability to predict the experimental value of the quantum system in a single experiment, not in our ability to predict some average value from an ensemble of identical experiments. For example, the fact that the average position of a quantum system is a well defined concept for an ensemble of identical experiments in the Orthodox theory does not imply that the Orthodox theory accepts that a quantum system has a well defined position in each experiment.    

We now move to demonstrate that intrinsic dynamical properties, i.e., property states in Modal quantum mechanics, can be linked to weak values.
For that let us note that in a given Modal theory, with a given \textit{property state} $|g^i(t)\rangle$, other properties $S_M(g)$, different from $G$, can also be considered to be real (independently of whether they are being measured or not) in a given Modal theory. The property $S_M(g)$ is linked to the \textit{guiding state} $|\psi(t) \rangle$ and the \textit{property state} $|g^i(t)\rangle$. At each time, in a particular $i$-th experiment, the value of the property $S_M(g)$ is given by $S_M(g^i(t))$. Then,  
the ensemble value of this property is given by:
\begin{equation}
\langle {S}\rangle_M = \lim_{N\rightarrow\infty} \frac{1}{N} \sum_{i=1}^N S_M(g^i(t)) = \int dg \;S_M(g)|\psi(g,t)|^2.
\label{modal2} 
\end{equation}
It is important to emphasize that the ensemble value of $\langle {S}\rangle_M$ is well-defined independently of the fact that the system is measured or not, because $S_M(g^i(t))$ is an \textit{ontic} variable.
In principle, it seems that the property $S_M(g)$ in \eref{modal2} can be arbitrary, but a necessary condition to accept $S_M(g)$ as a physically meaningful value linked to the property $S$ is that its expectation value is the Orthodox value $\langle {S}\rangle_M = \langle \hat S \rangle$, as it happens for the property $g^i(t)$ in \eref{modal1}. This last requirement leads to the following definition of $S_M(g)$:
\begin{equation}
    S_{M}(g)=\int dg'\frac{\psi^{*}(g,t)S(g,g')\psi(g',t)}{|\psi(g,t)|^{2}}
    \label{modal3}
\end{equation}
where $S(g,g')=\langle g |\hat S | g' \rangle$. When \eqref{modal3} is introduced in \eref{modal2}, it is then straightforward to realize that $\langle {S}\rangle_M = \langle \hat S \rangle$.
The central point of this paper is that the property $S_M(g)$ can be related to the so-called weak values by simply noting that \eref{modal3} can be rewritten as: 
\begin{eqnarray}
    S_{M}(g)=\frac{\langle g|\hat S|  \psi(t) \rangle}{\langle g|  \psi(t) \rangle} \equiv \;_{g}\langle \hat S \rangle_{\psi(t)},
    \label{modal5}
\end{eqnarray}
where we have used that $\int dg' |g'\rangle \langle g'|=\hat{I}$ in \eref{modal3}. The above identity states that the property $S_{M}(g)$ coincides with the weak value $_{g}\langle \hat S  \rangle_{\psi(t)}$ introduced by Aharonov, Albert and Vaidman in 1988 ~\cite{aharonov1988result}.

The quantity $S_M(g^i(t))$ is linked to two properties at the same time, so we can call it a \textit{dynamical} property. The last requirement to define $S_M(g^i(t))$ as an \text{intrinsic dynamical} property is the fact the quantum system has to be detached from any measuring apparatus. It is in this sense that we define intrinsic properties as \textit{context-free} properties. The condition for having the quantum system detached from the measuring apparatus in the evaluation of the intrinsic dynamical property or weak value in \eref{modal5}, is just ensuring a unitary evolution of the state $|\psi(t)\rangle$, from its preparation at the initial time until $t$, without including any additional degree of freedom different from the ones invoked at the initial time when defining the system (i.e. the system is a closed system during the whole evolution).    

As it will be discussed in the next section, the above result sets the grounds for understanding weak values as an experimental protocol for assessing the predictions of intrinsic dynamical properties of Modal quantum theories. But, more importantly, it reinforces the idea that understanding the (backaction-free) dynamics of quantum systems is possible by means of weak values.

\section{What are the implications of weak values and intrinsic properties being equivalent?}
\label{implications}
Since its first introduction in 1988 ~\cite{aharonov1988result}, there have been many attempts in the literature to find a fundamental interpretation of weak values, other than the simple result of a specific measurement procedure. Without abandoning the Orthodox quantum mechanics viewpoint, weak values have been given a number of different, often incompatible, interpretations~\cite{leggett1989comment,duck1989sense,zhu2011quantum,ferrie2014result,brodutch2015comment,piacentini2017determining,sinclair2019interpreting}; see Ref. \cite{matzkin2019weak} for a recent review on the difficulties to accommodate the weak values within the Orthodox ontology.

From the result in the previous section, we argue that the definition of a particular weak value has to be made under a particular Modal interpretations of quantum mechanics. \eref{modal5} shows that, once a particular weak value $_{g}\langle \hat S \rangle_{\psi(t)}$ and the corresponding intrinsic dynamical property $S_{M}(g)$ (linked to a particular Modal theory) are selected, the following ontological and empirical implications apply:\\

At the ontological level:
\begin{enumerate}[label=(\roman*)]
\setcounter{enumi}{0}
    \item The same ontological meaning of the intrinsic property $S_{M}$ can be given to the weak value $_{g}\langle \hat S \rangle_{\psi}$.
    
    \item If the intrinsic property $S_{M}$ cannot be given an Orthodox meaning, then $_{g}\langle \hat S \rangle_{\psi}$ cannot be given an Orthodox meaning either.
    
    \item The intrinsic property $S_{M}$ and the the weak value $_{g}\langle \hat S \rangle_{\psi}$ are context-free (i.e. defined with the system detached from any measuring apparatus). The discussion about contextuality of a property $S_{M}$ requires that the system interacts with a measuring apparatus, but then the property $S_{M}$ is no longer an intrinsic property.
\end{enumerate}

At the empirical level:

\begin{enumerate}[label=(\roman*)]
\setcounter{enumi}{3}
    \item Since $_{g}\langle \hat S \rangle_{\psi}$ can be experimentally evaluated, the Modal property $S_{M}$ is experimentally accessible too.    
    \item Since $_{g}\langle \hat S \rangle_{\psi}$ is in general a complex number, the Modal property $S_{M}$ takes a complex value too.
\end{enumerate}
That is, the identity in \eref{modal5} defines intrinsic Modal properties as weak values and in turn provides a clear-cut physical meaning of weak values in terms of intrinsic dynamical properties.

Let us discuss point (i) above. We want to clarify that conclusions arising from \eref{modal5} are also compatible, if wanted, with some opinions in the literature where weak values are understood as just mathematical transition amplitudes, without any direct physical implication concerning the definition of a property. This interpretation would simply require disregarding the particular Modal theory that gives support to the weak value through $S_M(g)$. See a clear example in \cite{App_B} where the weak value of spin is not accepted as a valid intrinsic property in one Modal theory, but it is accepted in another.

Regarding point (ii), in the particular case where $\hat S$ and $\hat G$ commute, $\langle g |\hat S | g' \rangle=s \delta(g-g')$ and \eref{modal3} reduces to $S_M(g)=s$, where $s$ is an eigenvalue of $\hat S$. Now, the Modal property $S_M$ is a well defined property in Orthodox quantum mechanics as well, as it fulfills the eigenstate-eigenvalue link. However, in more general cases where $\hat S$ and $\hat G$ are non-commuting operators, understanding the value $S_M(g)$ as a property of a quantum system from the perspective of Orthodox quantum mechanics is prevented by the the eigenvalue-eigenstate link.

In point (iii) we emphasize that there is no contradiction in arguing that intrinsic properties and weak values are context-free and quantum mechanics is contextual since we are talking about different things. The argument that quantum mechanics is contextual refers to the measured properties of a quantum system, while the intrinsc properties refers, by contsruction, to properties that are not being measured. Notice that the ontic property $S_M(g^i(t))$ of a Modal theory is not always an intrinsic dynamical property. When a measuring apparatus interacts with the system, the ontic property  $S_M(g^i(t))$ can change its value and hence becomes a contextual ontic property but, then, the property $S_M(g^i(t))$ is no longer an intrinsic dynamical property (which requires to be detached from a measuring apparatus). In other words, the adjectives context-free and non-contextual are different. The first refers to a system without (measuring) context, while the second refers to a system with a (measuring) context. The discussion about contextuality, originated from the Kochen-Specker theorem~\cite{Kochen1975}, is very relevant in the Orhtodox theory since it has direct implications on what properties can be considered simultaneously real in that theory (via the eigenstate-eigenvalue-link). On the contrary, in Modal theories, the reality of a property $g^i(t)$, or $S_M(g^i(t))$, is not linked to any measurement process. In Modal theories, the discussions about contextuality or non-contextuality only deals with how much experimental perturbation appears on the system due to the interaction with the measuring apparatus. See a more elaborated example of the differences between measured components of the spin and weak values of the components of the spin in \cite{App_B}.

Moving to point (iv), it is well known that weak values $_{g}\langle \hat S \rangle_{\psi}$ can be accessed experimentally~\cite{lundeen2009experimental,lundeen2011direct,hariri2019experimental}. From the identity in \eref{modal5}, intrinsic properties can be thus identically measured. It is important here to distinguish between measuring two properties in a single experiment (which has implications for the Orthodox reality of such properties), and measuring two properties in an ensemble of identically-prepared experiments from an average over different data. Weak values (for example, the Bohmian velocity) belong to this second type, but they are \textit{genuine} measurements ~\cite{durr2009weak}. The most relevant feature of the experimental determination of the weak value is that of providing information of two properties at the same time in the absence of measurement  disturbance (back-action). It is not possible to get a single system linked to a measuring apparatus giving empirical data free from back-action when two properties are measured in a single experiment (see~\cite{App_A}). But it is possible to get an ensemble of identically-prepared systems (each one suffering from back-action when two properties are measured) that provides a post-processed data (in a way that the different back-actions compensate) identical to the intrinsic dynamical property of a single system (see~\cite{app_C}).

As stated in (v) above, the meaning given to the real and imaginary parts of the weak value $_{g}\langle \hat S \rangle_{\psi(t)}$ must be the same as given to the real and imaginary parts of the intrinsic property $S_{M}(g)$. A physical interpretation of the weak value can be easily given from the perspective of the measuring apparatus as a shift in the pointer’s mean position and mean momentum~\cite{jozsa2007complex}. From the perspective of the quantum system, however, an interpretation of weak values is not so straightforward. Unlike the real part of the weak value, which can be shown to provide direct information on the dynamics in time of a given observable in the limit of zero measurement disturbance (i.e., an intrinsic property), the imaginary part of the weak value does not necessarily provide information pertaining to the observable being measured~\cite{leggett1989comment,jozsa2007complex,dressel2012significance}. Importantly, the imaginary parts of $S_{M}(g)$ and $_{g}\langle \hat S \rangle_{\psi(t)}$ do not play any role in the evaluation of the ensemble value $\langle {S}\rangle$ that motivated our definition of $S_M(g)$ in \eref{modal3}. That is, the weighted sum over the imaginary part of $S_{M}(g)$ and $_{g}\langle \hat S\rangle_{\psi(t)}$ in \eref{modal2} vanishes because $\langle {S}\rangle$ is a real number for any Hermitian operator $\hat S$. 

A more in deep discussion about the physical meaning of the imaginary part of the intrinsic properties and weak values is, however, beyond the scope of this paper. Hereafter we will  focus on the discussion of their real part. For that, we introduce the real part of intrinsic Modal properties as:
\begin{equation}\label{modal6}
    S_{M}^{Re}(g) = \text{Re}\left[S_{M}(g)\right] = \text{Re} \left[ \;_{g}\langle \hat S \rangle_{\psi(t)}\right].
\end{equation}
Using the above expression, the probability density distribution of the different values of the intrinsic property $S_{M}^\text{Re}(g)$ can be then easily written as:
\begin{equation}
P_M(s,g,t)=\lim_{N\rightarrow\infty}\frac{1}{N}\sum_{i=1}^{N}\delta[s-S_M^{Re}(g^i(t))].
\label{p(s)}
\end{equation}
 
The above expression highlights one of the most important advantages of working with intrinsic properties. When properly integrated, expression \eqref{p(s)} counts how many times the following identities $s=S_M^{Re}(g^i(t))$ and $g=g^i(t)$ are satisfied simultaneously. By construction, the number of coincidences can be zero or any other positive number. But, the probability $P_M(s,g,t)$ cannot be negative. This is not always true when computing probability with other more Orthodox tools. For example, the Wigner function distribution, whch provides simultaneous information on the position $x=g$ and momentum $p=s$, has negative values \cite{zhan2016unphysical}. It is in this sense that the Wigner distribution function is a quasi-probability distribution to account for its negatives values. The probability distribution computed from $P_M(s,g,t)$ are always complete positive \cite{colomes2017quantum}

\section{Intrinsic Bohmian dynamics as local-in-position weak values}
\label{examples_unmeasured}

As we have shown in the previous section, weak values have to find ontological support in Modal theories. For example, if we are interested in local-in-momentum weak values $\langle p|\hat S \psi\rangle/\langle p|\psi\rangle$, then the property state of such Modal theory has to be the momentum. A momentum-based Modal theory can be developed, for instance, by considering the state $\langle p|\psi\rangle$ and rewriting the Schr\"odinger equation in terms of $\langle p|\psi\rangle$ (instead of $\langle x|\psi\rangle$). A Hamilton-Jacobi decomposition can be then used to find an equation of motion for $p(t)$, which is defined as an intrinsic property of the quantum system~\cite{bonilla2020bohmian}. As discussed in the previous section, such a Modal theory would provide a full ontological meaning for local-in-momentum weak values (but not for local-in-position weak values which are defined in the original Bohmian theory).

It seems reasonable to demand that, for a Modal theory to be consistent, properties $S_M(g)$ that have a \textit{natural} connection with $G$ do also have to have an ontological reality. Let us provide a practical example of what we mean by \textit{natural} connection. In Bohmian mechanics~\cite{bohm52,holland1995quantum,pladevall2019applied,durr13}, the \textit{property state} for a given experiment $|g^i(t)\rangle \equiv |x^i(t)\rangle$ specifies the position of all particles $g^i(t) \equiv x^i(t)$ at all times (without the need to introduce the measuring apparatus)~\cite{bub1996modal}. 
The Bohmian version of \eref{modal6} reads:
\begin{equation}
S_B(x) =\mbox{Re}\left[ _{x}\langle \hat S \rangle_{\psi(t)} \right],
\label{sb_bis}
\end{equation}
so, the real part of local-in-position weak values are identical to intrinsic Bohmian properties. Now, if we postulate that the position is an intrinsic property, then, the temporal derivative of such intrinsic property, i.e., $d x^i(t)/dt$, must be also a well-defined property at all times. Including the mass $m$ of the quantum system, the Bohmian property of the momentum defined at all times (with or without measurement) has to be:
\begin{equation}
P_B(x^i(t)) \equiv m \; \frac{d x^i(t)}{dt}
\label{momentum}
\end{equation}
But, is this particular expression of the Bohmian momentum (\ref{momentum}), coming from a \textit{natural} deduction from $x^i(t)$, compatible with the general expression (\ref{modal3})? Will  the Bohmian momentum (\ref{momentum}) satisfy $\langle {P}\rangle_B = \langle \hat P \rangle$? By introducing the momentum operator $\hat P$ in the position representation given by $\langle x|\hat P|x'\rangle=-i\hbar \frac{\partial}{\partial x} \delta(x-x')$ into \eqref{modal3}, it is well-known that expression (\ref{momentum}) is exactly reproduced ~\cite{holland1995quantum, pladevall2019applied} showing a \textit{natural} consistency of the Bohmian theory. 

The discussion about the internal consistencies of other Modal theories (see ~\cite{dieks2007probability}) is however not in the scope of this paper, and in the rest of the paper we will focus on showing the physical soundness and utility of the intrinsic Bohmian properties linked to local-in-position weak values. As it will be shown, the knowledge of the position of a particle without linking its value to the measuring pointer provides some advantageous framework to answer some puzzling questions for example the time spent by a quantum particle to go from one place to another. This information is required, e.g., to evaluate transit and dwell times, but it is also necessary to define spatial distributions of some relevant quantities such as the (thermodynamic) work. Identically, when dealing with high frequency quantum electron transport, the existence of the position of particles independently of the measuring apparatus allows to know what is the contribution of electrons (at different positions) to the total, particle plus displacement, electrical current.


\subsection{The quantum dwell time}
Measuring the time spent by a particle within a particular region $\vec a<\vec r<\vec b$ requires measuring the time $t_1$ at which 
the particle enters that region and later, the time $t_2$ at which the particle leaves it. 
As we have already seen, the measurement of the position of the particle implies the perturbation of the state of the system in most general circumstances.
Thus, any subsequent measurement of the position is generally influenced by the first measurement.  
In spite of its controversial definition in Orthodox quantum theory, the concept of dwell time is necessary, for example, to evaluate the maximum working frequency of state-of-the-art transistors and hence the performance of modern computers \cite{zhan2017limitations}. 
In this respect, it is important to notice that when using the information of the dwell time in the evaluation of the performance of computers, there are no position detectors at the two ends ($\vec a$ and $\vec b$) of the active region of a transistor~\cite{tunnel_time, tun_review}.

The \textit{unperturbed} value of the dwell time can be easily computed from \textit{intrinsic} Bohmian trajectories $\vec r^i(t)$. Again, for simplicity, we only consider one electron inside the active region in each experiment. The expectation value of the unperturbed dwell time can be defined as:
\begin{equation}
\centering
\tau_{D}=\lim_{N\rightarrow\infty}\frac{1}{N}\sum _{i=1}^{N}\tau^i ,
\label{tun1}
\end{equation}
where $\tau^i$ is defined as the time spent by the (intrinsic) $i^{\rm th}$ Bohmian trajectory inside the region $\vec a<\vec r<\vec b$, i.e.: 
\begin{equation}
\centering
\tau^i=\int_{0}^{\infty}dt\Theta[\vec r^i(t)-\vec a]\Theta[\vec b-\vec r^i(t)],
\label{tun2}
\end{equation}
where $\Theta$ is the unit step function. The above expression can be rewritten as
\begin{equation}
\centering
\tau_{D}=\lim_{N\rightarrow\infty}\frac{1}{N}\sum _{i=1}^{N}\int_{0}^{\infty}dt \int_{\vec a}^{\vec b} \delta[\vec r-\vec r^i(t)] d\vec r.
\label{tun3}
\end{equation}
Using \eqref{velo1} and \eref{weak_current}, $\vec r^i(t)$ can be defined in terms of weak values as:
\begin{equation}
    \vec r^i(t) = \vec r^i(0) + \frac{1}{m}\int_0^t dt'\text{Re}\left[_{\vec r^i(t')}\langle \hat{P}_x \rangle_{\psi(t')}\right],
    \label{tun_P}
\end{equation}
with $\vec r^i(0)$ as the initial position of the trajectory in $i^{\rm}$-th experiment. Making use of the quantum equilibrium condition \cite{Oriols12} in \eref{tun3}, we get the well-know expression
\begin{equation}
\centering
\tau_{D}=\int_{0}^{\infty}dt \int_{\vec a}^{\vec b} |\psi(\vec r,t)|^2 d\vec r ,
\label{tun4}
\end{equation}
which is certainly an unperturbed property of the quantum system as there is no
contamination from the measuring apparatus. Notice that the experimental validation of the above arguments requires to know the intrinsic Bohmian trajectories, which in turn can be reconstructed from local-in-position weak values of the momentum~\cite{wiseman2007grounding}. 

Equation \eref{tun_P} provides a way to compute the intrinsic value of the dwell time from local-in-position weak values of the momentum operator $\hat{P}$. Alternatively, we can compute this time from a definition of the dwell time operator $\hat{D}$ as follows,
\begin{eqnarray}
	\tau_{D}&=&\int_{0}^{\infty} d t \int_{\vec a}^{\vec b}|\psi(\vec r, t)|^{2} d \vec r\nonumber\\
	&=&\int_{0}^{\infty} d t \int_{\vec a}^{\vec b}\langle\psi(t) \mid \vec r\rangle\langle \vec r \mid \psi(t)\rangle d \vec r.
	\label{tun_alternate1}
\end{eqnarray}
Defining the operator $\hat{A}=\int_{\vec a}^{\vec b}|\vec r\rangle \langle \vec r|$, \eref{tun_alternate1} can be written as
\begin{equation}
	\tau_D=\int_{0}^{\infty} dt\langle \psi(t)|\hat{A}| \psi(t)\rangle=\left\langle\psi(0)\left|\int_{0}^{\infty}dt \hat{U}^{\dagger} \hat{A} \hat{U}\right| \psi(0)\right\rangle
	\label{tun_alternate2}
\end{equation}
where we have defined $|\psi(t)\rangle=\hat{U}(t)|\psi(0)\rangle$. Now we define the tunnelling time operator $\hat{D}=\int_{0}^{\infty} dt \hat{U}^{\dagger}(t) \hat{A} \hat{U}(t)$, which allows us to write \eref{tun_alternate2} as follows,
\begin{equation}
	\tau_D=\left\langle\psi(0)|\hat{D}| \psi(0)\right\rangle.
\end{equation}
We can easily represent the above expression in terms of weak values of the tunnelling time operator $\hat{D}$ (with post-selected state $|\vec{r}\rangle$ and the initial state $|\psi(0)\rangle$) by a simple transformation as follows,
\begin{eqnarray}
\tau_D&=&\int d \vec{r}\left\langle\psi(0) \mid \vec{r}\right\rangle\langle \vec{r}\;|\hat{D}| \psi(0)\rangle\nonumber\\
&=&\int d \vec{r}\;\langle\psi(0) \mid \vec{r}\rangle \; \langle\vec{r} \mid \psi(0) \rangle \frac{\langle \vec{r}\;|\hat{D}| \psi(0)\rangle}{\langle\vec{r} \mid \psi(0) \rangle}\nonumber\\
&=&\int d \vec{r}\;\left|\psi(\vec{r},0))\right|^{2}\;_{\vec{r}}\langle\hat{D}\rangle_{\psi(\vec{r},0)}\nonumber\\
&=&\lim_{N\rightarrow\infty}\frac{1}{N}\sum _{i=1}^{N} {_{\vec{r}^i(0)}}\langle\hat{D}\rangle_{\psi(\vec{r},0)},
\label{tun_alternate4}
\end{eqnarray}
where we have used the quantum equilibrium \cite{pladevall2019applied} expression $\left|\psi(\vec{r},0)\right|^{2}=\lim_{N\rightarrow\infty}\frac{1}{N}\sum _{i=1}^{N} \delta(\vec{r}-\vec{r}^{i}(0))$ in the last identity of \eref{tun_alternate4}. By comparing \eref{tun1} and \eref{tun2} to \eref{tun_alternate4}, it is easy to obtain the relation between the Bohmian intrinsic tunnelling time $\tau^i$ and the local-in-position weak value of the tunnelling time $_{\vec{r}^i(0)}\langle\hat{D}\rangle_{\psi(\vec{r},0)}$, i.e.:  
\begin{equation}
\tau^i = \;\; _{\vec{r}^i(0)}\langle\hat{D}\rangle_{\psi(\vec{r},0)}.
\label{tun_alternate5}
\end{equation}
Notice that, at $t=0$, we get $\hat U(0)$ equal to the unity, so that the evaluation of the weak value of the tunneling time without the time integral is $\int_{\vec a}^{\vec b} d\vec{r} \langle \vec r^i(0)|\vec r \rangle \langle \vec r| \psi(0) \rangle/(\langle \vec r| \psi(0) \rangle) = \int_{\vec a}^{\vec b} d\vec{r} \langle \vec r^i(0)|\vec r \rangle=\Theta[\vec r^i(0)-\vec a]\Theta[\vec b-\vec r^i(0)]$, which is exactly the expression of the intrinsic tunneling time in \eref{tun2} without the time integral. A similar argument can be applied at any other time $t$ by just redefining the initial time and evolving the trajectories and wave functions accordingly. \Eref{tun_alternate5} says that the Bohmian dwell time of the $i$-th particle is identical to the weak value of the dwell time operator associated to the $i$-th particle defined as the one whose initial position is $\vec{r}^i(0)$. Therefore, \eref{tun_alternate5} shows a deep connection between weak values and intrinsic dynamical properties. 

Certainly, there exist many Orthodox protocols to compute the dwell time \cite{tun4,tun20,tun22,tun80,tun24,tun25}.  
For example, one can make use of a physical clock to measure the time elapsed during the tunneling \cite{tun22, tun80, tun24, tun25}. 
Larmor precession was precisely introduced to measure the time associated with scattering events \cite{tun80,tun25}. 
In any case, what is essential here is that the scientific community has been persistent in looking for observables of dynamical properties whose expectation value is free from the contamination of the measuring apparatus.
This is exactly what \textit{intrinsic} properties defined in this paper are meant for.

\subsection{The quantum work distribution}

Quantum work is the basic ingredient in the development of quantum thermodynamics which is one of the most important topics in the field of open quantum systems. Quantum thermodynamics is essential in developing new quantum technologies such as quantum heat engines. It also plays a fundamental role in the consistency of the second law of thermodynamics in the quantum regime. However, there are many issues that are still being investigated, most notably related to the definition of work and heat. The problem is that these thermodynamic variables are not observables related to Hermitian (super)operators, but are trajectory (history) dependent \cite{vilar2008failure,gelbwaser2017thermodynamic,niedenzu2019concepts}.  This has culminated in the so-called "no-go" theorem that states that in fact there cannot exist a (super)operator for work that simultaneously satisfies all the physical properties required from it \cite{work58}.
This conclusion is based on three requirements to be fulfilled by what 
Acin {\it et al.} \cite{work58} define to be a properly defined positive definite work distribution. Namely:
\begin{enumerate}
    \item The work distribution is described by a positive operator valued measurement (POVM).
    \item For initial states that commute with the initial Hamiltonian the work distribution reduces to the two-point measurement (TPM) work distribution.
    \item The work distribution respects the change in the Hamiltonian expectation value.
\end{enumerate}

As indicated in the introduction, evaluating quantum work by means of a TMP measurement protocol of the energy will imply that the first measurement projects the initial state into an energy eigenstate, hence preventing the possibility of capturing any coherent evolution of the state. The alternative Orthodox protocols for the evaluation of the quantum work such as Gaussian measurements \cite{work66,work67}, weak measurements \cite{work54,work70}, collective measurements \cite{work58}, etc. all suffer from quantum contextuality, which provides as many different work definitions as measurement schemes exist. The problem appears due to the requirement of the Orthodox theory to include a measuring apparatus that in practice does not exist. In other words, we are not interested in the explicit measurement of work, but on using dynamical information of the quantum (sub)system in conjunction with quantum thermodynamic equations to compute, e.g.,
the temperature variation of a larger system involving a macroscopic thermodynamic environment.
We are thus seeking for an \textit{unperturbed} value of work. 

To define an unperturbed work distribution, we follow similar steps as in Refs. \cite{kobe,work}, where the reader can find a detailed derivation of quantum work based on Bohmian mechanics. Here, we will quickly move to the definition of intrinsic work to discuss in detail how this definition solves the above described puzzling situation.
We thus start by defining a single particle\cite{footnote3} wave function solution of the following Schr\"{o}dinger equation:
\begin{equation}
\centering
i\hbar \frac{\partial \psi(\vec r,t)}{\partial t}=\left(\frac{(-i\hbar \vec \nabla -q \vec A(\vec r,t) )^2}{2m}+qV(\vec r,t) \right) \psi(\vec r,t).
\label{work1}
\end{equation} 
 where $\vec r$ is defined as a vector in the ordinary three dimensional space, $\vec \nabla$ is the gradient operator and $-i\hbar \vec \nabla -q \vec A(\vec r,t)$ is the  canonical momentum with $A(\vec r,t)$ the electromagnetic vector potential. When the wave function is written in polar form as $
\psi(\vec r,t)=R(\vec r,t)\mbox{exp}\left(\frac{i\mathcal{S}(\vec r,t)}{\hbar}\right)$ where $R(\vec r,t)$ and $\mathcal{S}(\vec r,t)$ are the modulus and phase, respectively, the real part of \eref{work1} evaluated along the Bohmian trajectory $\vec r=\vec r^i(t)$ for the $i^{\rm th}$ experiment, gives us the following equation for the unperturbed power:  
\begin{flalign}
\frac{d \mathcal{E}(\vec r^i(t))}{d t} = \frac{d}{d t}\left (\frac {1}{2}m \vec v^2(\vec r^i(t),t)+ Q(\vec r^i(t),t) \right)\nonumber\\
= q \vec v(\vec r^i(t),t)  \vec E(\vec r^i(t),t) +\frac{\partial Q(\vec r^i(t),t)}{\partial t}.&
\label{work3}
\end{flalign}
Here $\mathcal{E}(\vec r^i(t),t)$ is the unperturbed energy of the system which according to the Hamilton-Jacobi equations is the energy of the configurations $\Vec{r}$ which is given by the time derivative of the argument of the wavefunction, i.e. $\mathcal{E}(\Vec{r},t)=-\partial \mathcal{S}(\Vec{r},t)/\partial t$, $v(\vec r^i(t),t)$ is the Bohmian velocity, $Q(\vec r^i(t),t)$ is the quantum potential and $\vec E(\vec r^i(t),t)$ is the electric field. While we have considered an external electromagnetic field interacting with the quantum system, 
no measuring apparatus is accounted for in \eref{work3}. 
Thus, from \eref{work3}, we can describe the unperturbed work represented by the wave function $\psi(\vec r,t)$ and the trajectory $\vec r^i(t)$, during the time interval $t_2-t_1$ by just 
subtracting the initial energy $\mathcal{E}(\vec r^i(t_1),t_1)$ from the final one $\mathcal{E}(\vec r^i(t_2),t_2)$. As we have already mentioned, this result corresponds to the single experiment labelled by the superscript $i$. Getting ensemble values of the work just requires repeating the previous procedure for different initial positions of the particles, according to the quantum equilibrium hypothesis~\cite{Oriols12}.\\

As stressed along the paper, the very crucial aspect of this unperturbed work is its measurability in the laboratory using the local-in-position weak values. This can be expressed as follows, 
\begin{equation}
\text{Re}\left(_{\vec r^i(t)}\langle \hat{\mathcal{E}} \rangle_{\psi(t)}\right) =\mbox{Re}\left(\frac{\langle \vec r^i(t)|\hat H|\psi(t)\rangle}{\langle \vec r^i(t)|\psi(t)\rangle}\right).
\label{weak_energy}
\end{equation}
Here we define ${\mathcal{\hat{E}}\equiv \hat{H}}$ which is just the usual nomenclature to define quantum work. The left hand side of \eref{weak_energy} corresponds to the weak value of the energy computed at a particular trajectory that has been obtained by post-selecting the position. Now using the Hamiltonian from \eref{work1} in \eref{weak_energy} and writing the wave function in the polar form to evaluate $\langle \vec r^i(t)|\hat H|\psi(t)\rangle$, we can rewrite \eref{weak_energy} as:
\begin{eqnarray}
\text{Re}\left(_{\vec r^i(t)}\langle \hat{\mathcal{E}} \rangle_{\psi(t)}\right)&=&\left[\frac{-\hbar^2}{2m}\frac{\nabla ^2R(\vec r,t)}{R(\vec r,t)}+\frac{(\vec \nabla \mathcal{S}(\vec r,t))^2}{2m}\right]_{r=r^i(t)}\nonumber\\
&=&Q(\vec r^i(t),t)+\frac {1}{2}m \vec v(\vec r^i(t),t)^2,
\end{eqnarray}
where $\nabla^2$ is the Laplacian operator. Thus, from \eref{work3}, it is straightforward to see  that,
\begin{equation}
\text{Re}\left(_{\vec r^i(t)}\langle \hat{\mathcal{E}} \rangle_{\psi(t)}\right)=  \mathcal{E}(\vec r^i(t),t).  
\end{equation}
We thus conclude that the intrinsic Bohmian energy is equal to the local-in-position weak value of the energy and hence that it can be, in principle, measured experimentally. Given a collection of weak values of the energy at times $t_1$ and $t_2$, one can then easily evaluate the quantum work in the $i^{\rm th}$ experiment as,
\begin{equation}
    W^i(t_2,t_1) = \text{Re}\left(_{\vec r^i(t_2)}\langle \hat{\mathcal{E}} \rangle_{\psi(t_2)}\right) -  \text{Re}\left(_{\vec r^i(t_1)}\langle \hat{\mathcal{E}} \rangle_{\psi(t_1)}\right).
    \label{work_iexp}
\end{equation}
The work distribution on the other hand can be given as follows,
\begin{equation}
    \mathcal{P}(w,t_2,t_1) = \lim_{N\rightarrow\infty}\frac{1}{N}\sum_{i=1}^N \delta({w-W^i(t_2,t_1)}).
    \label{work_distribution}
\end{equation}
From the work distribution in \eref{work_distribution} we can now evaluate the corresponding expectation value: 
\begin{eqnarray}
    \langle W(t_2,t_1) \rangle &=&\int dw\; w\;\mathcal{P}(w,t_2,t_1)\nonumber\\
    &=& \lim_{N\rightarrow\infty}\frac{1}{N}\sum_{i=1}^N W^i(t_2,t_1).
\end{eqnarray}
where $N$ is the total number of experiments considered $i=1,2,..,N$. 

The above Bohmian approach circumvents the problem of the unavoidable contextuality of quantum work in the Orthodox theory. The unperturbed "Bohmian work" fulfills the aforementioned last two requirements (ii and iii) for a properly defined work distribution and reduces to the known definitions in the appropriate limits. Furthermore, it circumvents the no-go theorem in~\cite{work58} because it cannot be associated with a POVM. Instead, the Bohmian work always defines a positive value for the work probability distribution and, moreover, describes a property that is free from any measuring context (as it does not depend on the backaction of the measuring apparatus). These are new conditions that should, in our opinion substitute (i) in \cite{work58}, i.e.:
\begin{enumerate}
    \item[(i.1)] The work definition must always lead to a positive-valued probability distribution.
    \item[(i.2)] The work distribution should not depend on the backaction of the measurement. It should be context-free.
\end{enumerate}

\subsection{The quantum noise at high frequencies}

Electron devices are meant to manipulate information in terms of scalar potentials  and output electrical currents.  
State-of-the-art electron devices are nowadays entering into the nanometer scale and operating at frequencies of hundreds of GHz. 
At such time and length scales, a full quantum treatment of the electrical current is mandatory. 
Furthermore, the total current at such frequencies is the sum of the conduction (flux of particles) plus the displacement (time-derivative of the electric field) components \cite{ramo,shockley,ZhenTED,zhen2}. 

At very high-frequencies (on the order of THz) the environment correlations are expected to decay on a time-scale comparable to the time-scale relevant for the system evolution. In this highly non-Markovian regime, a measurement at a given time within the active region of the device will certainly influence the dynamics of the electrons and hence the outcome of a subsequent measurement~\cite{e21121148}. That is, multi-time observables are not observables related to Hermitian operators, but, as it happens with the quantum work, are trajectory (history) dependent. 
Therefore, if one is interested in defining multi-time characteristics of these devices independently of the measuring apparatus, this should be done through the introduction of some sort of intrinsic electrical current. This idea is reinforced by the fact that it can be proven the electrical current in the device is very weakly perturbed by the electrons of the cables and surroundings and hence that measuring the electrical current at a macroscopic distance does not perturb appreciably electron dynamics in the active region~\cite{weak_current}.

To simplify the discussion, let us consider the \textit{unperturbed} value of the total electrical current for a single electron in the active region of the device (the extension to many electrons is conceptually straightforward). Given the Bohmian trajectory $\vec r^i(t)$ of the electron (which is obtained solving a transport equation such as in \eref{work1}), the total current generated on a surface $S$ of the active region of the device is:

\begin{equation}
I^i(t)=\int_{S} \vec J^i_c(\vec r,t) \cdot \mathrm d \vec s +\int_{S} \epsilon(\vec r,t)\frac{\mathrm d\vec E^i(\vec r,t)}{\mathrm d t}\cdot \mathrm d \vec s,
\label{i1}
\end{equation}
where $\epsilon(\vec{r},t)$ is the (inhomogeneous) electric permittivity, which can be time-dependent. The current (particle) density for the $i^{\rm th}$ experiment is given by $\vec{J}^i_c(\vec{r},t)=q \vec v(\vec r^i(t),t) \delta[\vec r-\vec r^i(t)]$ with $q$ the electron charge. The electric field $\vec E^i(\vec r,t)$ is the solution of the Gauss equation\cite{footnote4} with proper boundary conditions for a charge density given by $Q^i(\vec r,t)=q \delta[\vec r-\vec r^i(t)]$. In principle, the surface integral of the current density in \eref{i1} has a dependence on the particle position $\vec r^i(t)$. However, due to the conservation of the total current, it can be proven that for a two terminal device with a length $L$ much smaller than the lateral dimensions $W,H$, i.e., $L \ll W,H$, then the total current can be well approximated as~\cite{ramo,shockley,pellegrini}:
\begin{equation}
I^i(t)=\frac{q}{L}v_x^i(t),
\label{i4}
\end{equation}
where $v_x^i(t)$ is the velocity of the electron in the active region, i.e., $\frac{d}{dt} x^i(t)$~\cite{footnote5}. Outside the active region, the electron is screened and its contribution to the total current can be neglected. 

Using the definition of the Bohmian velocity $v_x(t)$~\cite{pladevall2019applied}, the intrinsic electrical current in \eref{i4} can be now expressed in terms of the local-in-position weak value of the momentum operator. 
\begin{equation}
v_x(t)=\frac{\hbar}{m}\text{Im}\left[\frac{\frac{\partial}{\partial x}\psi(\vec r,t)}{\psi(\vec r,t)}\right]=\frac{1}{m}\text{Re}\left[\frac{\langle \vec r|\hat P_x|\psi(t)\rangle}{\langle \vec r|\psi(t)\rangle}\right],
\label{velo1}
\end{equation}
where we have used $\langle \vec r|\hat P_x|\psi(t)=-i\hbar\frac{\partial}{\partial x}\psi(\vec r,t)$. Evaluating the velocity in \eref{velo1} for a particular trajectory $\vec r=\vec r^i(t)$ we finally get the current for the $i^{\rm th}$ experiment from \eref{i4} as follows:
\begin{equation}
I^i(t)=\frac{q}{mL}\text{Re}\left[\frac{\langle \vec r^i(t)|\hat P_x|\psi(t)\rangle}{\langle \vec r^i(t)|\psi(t)\rangle}\right]=\frac{q}{mL}\text{Re}\left[_{\vec r^i(t)}\langle \hat{P}_x \rangle_{\psi(t)}\right]. 
\label{weak_current}
\end{equation}
Thus the unperturbed current is equivalent to the local-in-position weak measurement of the momentum operator $\hat P_x $ in the transport direction ($x$), where $\langle \vec r|\psi(t) \rangle =\psi(\vec r,t)$ is the wave function of the electron in the active region and $r^i(t)$ specifies the position where the local-in-position weak value of the momentum is to be evaluated. From the information of an ensemble of experiments, it is then easy to evaluate the ensemble value of the current as:
\begin{equation}
\langle I(t) \rangle=\lim_{N\rightarrow\infty}\frac{1}{N}\sum_{i=1}^N  \frac{q}{mL}\text{Re}\left[_{\vec r^i(t)}\langle \hat{P}_x \rangle_{\psi(t)}\right].
\end{equation}

The intrinsic current in \eqref{weak_current} can be now used to define (dynamical) properties that depend on multiple-time observations without introducing the perturbation that the first measurement produces on the subsequent dynamics of the electron. 
Consider, for example, the power spectral density of the electrical current, which can be calculated through the Fourier transform of the current-current correlations. This quantity requires the knowledge of the current at two different times and hence its Orthodox definition would always be contextual (as the first measurement of the current would affect the subsequent dynamics of the electron). Alternatively, an intrinsic power spectral density can be easily defined relying on the intrinsic current in \eqref{weak_current}~\cite{BITLLES1,BITLLES2,BITLLES3,BITLLES5,BITLLES6,BITLLES7,BITLLES8}:
\begin{eqnarray}\label{PSD}
    \mbox{PSD}(\omega) = \lim_{N\rightarrow\infty}\frac{1}{N}\sum_{i=1}^N \left(\frac{q}{mL}\right)^2  \int_{-\infty}^{\infty} d\tau e^{-i\omega\tau}...\nonumber\\ ...\; \text{Re}\left[_{\vec r^i(t_2)}\langle \hat{P}_x \rangle_{\psi(t_2)}\right]\text{Re}\left[_{\vec r^i(t_1)}\langle \hat{P}_x \rangle_{\psi(t_1)}\right],
\end{eqnarray}
where we have assumed that we are dealing with a wide-sense stationary process where the correlation depends only on the time difference $\tau=t_2-t_1$. 
Therefore, the intrinsic (backaction free) power spectral density (as well as any other intrinsic dynamical property) not only can be easily defined, but it can be measured using weak values.

\section{Conclusions:} 
\label{conclusion}

In this paper we have introduced the concept of intrinsic properties. Specifically, we have defined an intrinsic property as a property of a quantum system whose existence is independent of the measuring apparatus and hence that provides information about the unperturbed (backaction free) quantum dynamics. These intrinsic properties cannot be defined within the eigenvalue-eigenstate link of Orthodox quantum mechanics (unless the system is in an eigenstate of the associated property operator). Contrarily, intrinsic properties arise naturally in the context of Modal quantum theories (e.g., Bohmian mechanics), where the ontology of Orthodox quantum mechanics is augmented by introducing property states.

We have shown the equivalence between intrinsic properties and weak values. Such equivalence suggests that any attempt to provide weak values with a physical meaning is at the same time an effort to give physical meaning to the intrinsic properties of Modal theories (and vice versa). Contrarily, it seems not possible to physically interpret weak values (apart from being a mathematical transition amplitude) within the ontology of Orthodox quantum mechanics, as intrinsic properties have no meaning there. Notice that there is no reasons to expect that all weak values have the same ontological status, in the same way as not all measurable properties have the same ontological status in a given Modal theory. Thus, it is possible to discredit most of weak values as just the result of juggling with experimental data (without any physical significance), but at the same time admitting that few weak values have physical significance supported by one particular Modal theory.      

Intrinsic properties, or equivalently weak values, are, to a good approximation, accessible experimentally. This is a relevant statement, as we are persistently looking for dynamical properties of quantum systems whose expectation value is free from quantum backaction~\cite{tun_review,baumer2018}. In particular, we have focused on local-in-position weak values or, equivalently, intrinsic Bohmian properties. We have shown that Bohmian trajectories, evaluated through weak values, allow to define and measure dwell times, quantum work statistics, or the power spectral density of current-current correlation functions. All these physical properties are associated to non-commuting operators.

Finally, we clarify that, while we have illustrated the soundness of intrinsic properties using three intrinsic Bohmian properties, the results in this paper do not put the Bohmian theory (i.e. the positions of quantum particles) in a privileged status. Our paper does only indicate that if Bohmian theory is not considered to provide a correct description of the quantum world, the same conclusion can be said about local-in-position weak values obtained in the laboratory. Equivalently, if local-in-position weak values are assumed to have a physical meaning, the same can be stated about the Bohmian theory. The same precise arguments apply to any other Modal theory and its associated weak values. Some experimental and theoretical examples of other Modal theories (i.e. with an ontological descriptions different from the Bohmian theory, while still yielding experimental predictions identical to those of the Orthodox quantum theory) can be found in Refs.~\cite{B-theories,mori2015quantum,matzkin2012observing,withers2015bilocal}. Our paper also emphasizes that looking for a meaning of weak values without linking them to a given ontology (or mixing ontologies in a type of bipolarity that wants a reality independent of the measurement, but at the same time rejects it) is not the correct path to understand weak values.

\section{Acknowledgements}
The authors are grateful to Alexandre Matzkin, Albert Sol\'e  and Travis Norsen for valuable discussions. D.P. and X.O acknowledge financial support from Spain's Ministerio de Ciencia, Innovaci\'on y Universidades under Grant No. RTI2018-097876-B-C21 (MCIU/AEI/FEDER, UE), the Generalitat de Catalunya and FEDER for the project QUANTUMCAT 001-P-001644, the European Union's Horizon 2020 research and innovation programme under grant agreement No 881603  GrapheneCore3 and under the Marie Skodowska-Curie grant agreement No 765426 (TeraApps). G.A. acknowledges financial support from the European Unions Horizon 2020 research and innovation programme under the Marie Skodowska-Curie Grant Agreement No. 752822, the Spanish Ministerio de Econom\'ia y Competitividad (Project No. CTQ2016-76423-P), and the Generalitat de Catalunya (Project No. 2017 SGR 348). T.A-N. and R.S. have been supported in part by the Academy of Finland through its QFT Center of Excellence Program grant (No. 312298).\\\\

\bibliography{refs}

\begin{thebibliography}{91}
\bibitem[{App({\natexlab{a}})}]{App_A}
\bibinfo{note}{A detailed proof of this statement can be found in Appendix A.}
\bibitem[{foo({\natexlab{a}})}]{footnote_new1}
\bibinfo{note}{The adjective \textit{intrinsic} is to be understood as
  'belonging only to the quantum system', without influence from the
  environment (here the measuring apparatus).}
\bibitem[{foo({\natexlab{b}})}]{footnote_new2}
\bibinfo{note}{Although there is a number of, slightly different,
  interpretations that fit under the umbrella of what is commonly called
  Orthodox quantum mechanics, we here refer to the standard presentation in
  textbooks, i.e.: (i) A quantum system at time $t$ is represented by a vector
  $|\psi(t)\rangle$ in a Hilbert space H. (ii) Properties of the system are
  represented by projectors on H. (iii) If and only if $|\psi(t)\rangle$ is an
  eigenstate of some projector, the system possesses the property associated
  with that projector; otherwise the value is ‘indefinite’ or
  ‘indeterminate’ or somesuch. (The ‘eigenstate-eigenvalue link’). (iv)
  A measurement of some property associated with projector $\hat P$ will find
  that property to be possessed by the system with probability $\langle\Psi|
  \hat P | \Psi\rangle$.}
\bibitem[{foo({\natexlab{c}})}]{footnote1}
\bibinfo{note}{{Many renowned physicists expressed their dissatisfaction
  regarding this issue. For example, Einstein wanted to discredit the
  ontological assumption that the measurement creates the position of a quantum
  object when he mentioned "Do you think the moon is not there when nobody is
  looking?". Similarly, Feynman once mentioned that "if a tree falls in a
  forest and there is nobody there to hear it, does it make noise?" to
  emphasize the difficulties to accept that reality of a physical process needs
  some type of observation to take place.}}
\bibitem[{foo({\natexlab{d}})}]{footnote_new3}
\bibinfo{note}{Modal theories may be compactly defined according to the
  following postulates: (i) A quantum system is represented by a guiding state
  $|\psi(t)\rangle$ plus a property state $|g^i(t)\rangle$ (for a given $i-$th
  experiment at time $t$) in a Hilbert space H. (ii) Properties of the system
  are represented by projectors on H. (iii) The system always possesses a well
  defined value of the property $G$ that is given by the value $g^i(t)$ of the
  property state. (iv) A measurement of some property associated with projector
  $\hat P$ will find that property to be possessed by the system with
  probability $\langle\psi(t)| \hat P | \psi(t)\rangle$.}
\bibitem[{App({\natexlab{b}})}]{App_B}
\bibinfo{note}{A detailed proof of this statement can be found in Appendix B.}
\bibitem[{app()}]{app_C}
\bibinfo{note}{A detailed proof of this statement can be found in Appendix C.}
\bibitem[{foo({\natexlab{e}})}]{footnote3}
\bibinfo{note}{A many-body treatment will not add any new insight on the
  discussion.}
\bibitem[{foo({\natexlab{f}})}]{footnote4}
\bibinfo{note}{Notice that this electric field $\vec E^i(\vec r,t)$ in
  \eref{i1} is different from the electric field $\vec E(\vec r,t)$ used in
  \eref{work3} in the previous section. Here, the electron located at $\vec
  r=\vec r^i(t)$ generates an electric field $\vec E^i(\vec r,t)$ which leads
  to a density current everywhere inside the device. In \eref{work3}, on the
  other hand, $\vec E(\vec r,t)$ was an external electric field without
  self-consistency with the electric charge of the electron considered.}
\bibitem[{foo({\natexlab{g}})}]{footnote5}
\bibinfo{note}{Typically, hundreds of electrons have to be considered in a
  realistic device and the total current in the $i$-th experiment can be
  computed by just adding the contribution of each electrons $I^i
  (t)=\sum_{n=1}^{N_e} I^i_n(t)$ with $N_e$ the number of electrons inside the
  volume $L\times W \times H$ at time $t$.}
\end{thebibliography}

\appendix
 
\end{document}